\documentclass[]{spie}  %>>> use for US letter paper
\usepackage{amsmath,amsfonts,amssymb}
\usepackage{graphicx}
\usepackage[colorlinks=true, allcolors=blue]{hyperref}
\usepackage{aasmacros}

\title{SynIM: a high-performance GPU-accelerated Python library for synthetic interaction and tomographic reconstruction matrices in next-generation adaptive optics}

\author[a,b]{Guido Agapito}
\author[a,b]{Fabio Rossi}
\author[a,b]{Alfio Puglisi}
\affil[a]{INAF -- Osservatorio Astrofisico di Arcetri, Largo E. Fermi 5, 50125, Firenze, Italy}
\affil[b]{ADaptive Optics National laboratory in Italy (ADONI)}

\authorinfo{Further author information: (Send correspondence to G.A.)\\G.A.: E-mail: guido.agapito@inaf.it}

\begin{document} 
\maketitle

\begin{abstract}
Next-generation Adaptive Optics (AO) systems for 8–40m class telescopes, such as MORFEO (ELT) and MAVIS (VLT), demand high calibration accuracy. Controlling thousands of actuators makes experimental calibration unfeasible due to daytime overheads, environmental disturbances, and opto-mechanical aberrations. Consequently, model-based (synthetic) calibration has become the mandatory standard. 

We present SynIM, an open-source Python library designed for computing high-fidelity synthetic Interaction, Projection, and Covariance Matrices. SynIM leverages GPU acceleration via CuPy to handle the massive dimensionality of high-order systems. A core innovation is its handling of spatial geometry via composite affine transformations and absolute sub-pixel grid alignment. By merging DM and WFS shifts, rotations, and magnifications into a single operation, SynIM minimizes interpolation artifacts. 

SynIM introduces an optimized numerical derivative engine for slope computation that mathematically aligns spatial grids at the sub-pixel level, closely mimicking the physical behavior of Shack-Hartmann sensors. It outperforms geometric estimators like the G-tilt telescoping sum at high spatial frequencies, while yielding a substantial computational speed-up. Crucially, end-to-end MCAO simulations demonstrate that reconstructors built with SynIM deliver closed-loop AO performance practically equivalent to full physical optics models. 

SynIM natively supports SCAO, GLAO, MCAO, and LTAO configurations. It features optimized multi-WFS batch processing, modules for MMSE tomographic reconstructors, and full compatibility with SPRINT for online tracking. Currently driving the design and operational strategies for MORFEO, MAVIS, AOF, KAPA, and WST, SynIM stands as an essential tool for next-generation AO calibration.
\end{abstract}

% Include a list of keywords after the abstract 

\section{INTRODUCTION}
\label{sec:intro}

The advent of Extremely Large Telescopes (ELTs) and the continuous upgrade of 8-10m class facilities have pushed Adaptive Optics (AO) into a new regime of complexity. Instruments like MORFEO for the ELT \cite{2021Msngr.182...13C,BusoniAO4ELT8} and MAVIS for the VLT \cite{2021Msngr.185....7R} rely on thousands of controlled actuators distributed across multiple Deformable Mirrors (DMs), guided by a constellation of Laser and Natural Guide Star (LGS/NGS) Wavefront Sensors (WFSs).

Historically, the calibration of AO systems relied on empirical measurements of the Interaction Matrix (IM), obtained by sequentially poking DM actuators and recording the WFS responses \cite{1990SPIE.1271...63B, 1991A&A...250..280R}. However, for next-generation facilities, this approach is fundamentally limited. The sheer scale of the system results in unacceptable daytime calibration overheads. Furthermore, empirical measurements are inevitably corrupted by dynamic environmental disturbances (e.g., dome seeing), opto-mechanical aberrations, and hysteresis, which degrade the matrix quality and ultimately limit the achievable Strehl ratio.

To overcome these physical and operational bottlenecks, Model-Based (synthetic) calibration has emerged as the mandatory standard for future astronomical facilities. A key conceptual milestone in this evolution was the Learn and Apply tomographic algorithm \cite{2010JOSAA..27A.253V}, which demonstrated the power of coupling empirical telemetry with analytical optical models. Building on these foundational principles, pseudo-synthetic calibration techniques were successfully deployed on the VLT Adaptive Optics Facility (AOF) \cite{2012SPIE.8447E..2DK, 2018SPIE10703E..1GO}, and subsequently extended to handle the non-linearities of systems equipped with Pyramid wavefront sensors \cite{2018MNRAS.481.2829H}. Today, the use of synthetically generated matrices is expanding across major observatories, driving recent hybrid calibration frameworks for the MMT and Keck AO upgrades \cite{10.1117/12.3020489, 10.1117/12.3017853,ConodAO4ELT8}. Ultimately, by computing the interaction and reconstruction matrices synthetically from high-fidelity physical models of the influence functions and optical geometry, it is possible to achieve a calibration free from measurement noise.

In this framework, we introduce SynIM (Synthetic Interaction Matrix)\footnote{The SynIM source code is publicly available at \url{https://github.com/ArcetriAdaptiveOptics/SynIM}, with full documentation hosted at \url{https://synim.readthedocs.io/}.}, a high-performance, open-source Python library designed to generate these calibration products.
Developed by the core team behind the PASSATA\cite{2016SPIE.9909E..7EA} and SPECULA\cite{specula2026} simulation frameworks, SynIM bridges the gap between system design and real-time operation. It natively computes synthetic interaction matrices, atmospheric covariance matrices, and tomographic projection operators for SCAO, GLAO, MCAO, and LTAO architectures.

To handle the immense computational load of ELT-scale matrices, SynIM is built from the ground up to leverage GPU acceleration via CuPy\cite{cupy_learningsys2017}, featuring smart memory caching and optimized batch processing for multi-WFS configurations. A critical architectural choice in SynIM is the use of composite affine transformations and sub-pixel grid alignment. Rather than sequentially resampling phase screens to account for DM-to-WFS misalignments, SynIM mathematically combines rotations, translations, and magnifications into a single transformation matrix, interpolating the high-resolution grid exactly once. 

This geometric rigor proved to be decisive for the slope computation engine. The computation of wavefront gradients on discrete grids is notoriously sensitive to sub-pixel misalignments. While analytical approaches like the telescoping sum offer high resilience against these discrete grid artifacts, SynIM demonstrates that enforcing strict sub-pixel spatial alignment fully enables numerical derivatives, allowing them to achieve high diffractive fidelity and computational speed.

This paper outlines the architecture of SynIM, detailing its advanced slope extraction algorithms, geometric alignment, tomographic MMSE reconstructor generation, and performance optimizations. We also present case studies demonstrating its critical role in the ongoing development and calibration of MORFEO, MAVIS, and other leading AO facilities.

\section{SYNIM ARCHITECTURE AND ALGORITHMS}
\label{sec:architecture}

The core philosophy of SynIM is to generate synthetic calibration matrices with strict mathematical rigor, ensuring that the numerical representation closely matches the physical behavior of the AO system. This requires handling high-resolution grids, complex composite rotations, and edge discontinuities efficiently. SynIM relies on a CuPy-based GPU backend\cite{cupy_learningsys2017}, enabling massive parallelization for ELT-scale computations while maintaining an automatic fallback to standard CPU processing via NumPy/SciPy if GPU memory is insufficient.

\subsection{Geometric Mapping and Single-Step Interpolation}
\label{subsec:geometry}

A typical MCAO or LTAO system presents a complex optical geometry: DMs and WFSs may have independent rotations, continuous sub-pixel shifts, magnification factors, and anamorphic distortions. Furthermore, LGS systems introduce altitude-dependent perspective scalings. 

\begin{figure}[ht]
    \centering
    \includegraphics[width=0.9\linewidth]{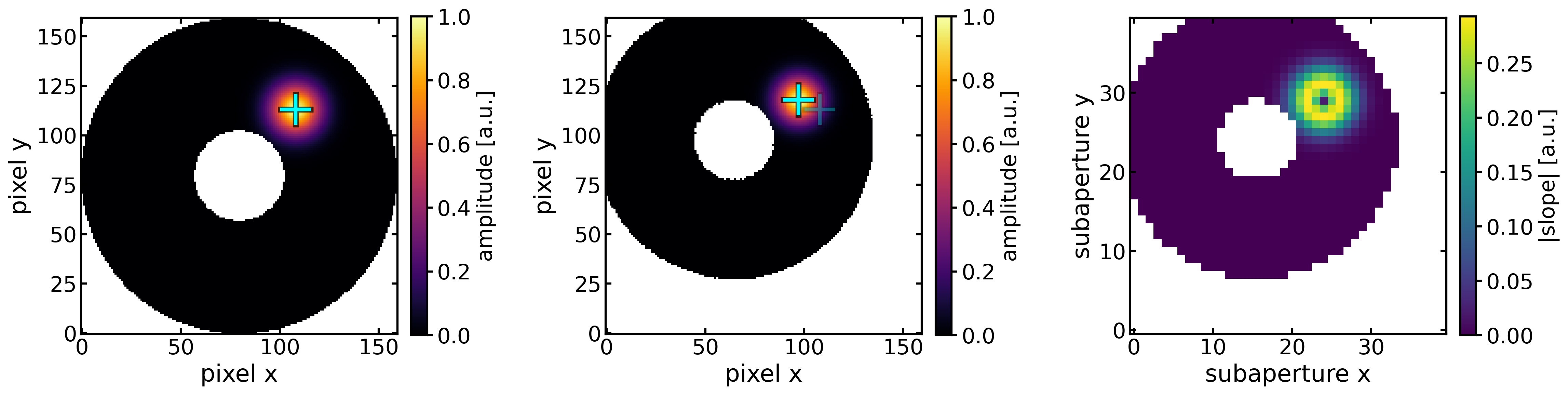}
    \caption{Visual representation of the SynIM geometric pipeline. The high-resolution influence function is mapped onto the WFS subaperture grid using a single composite affine transformation, ensuring sub-pixel alignment, and the resulting slope pattern (right).}
    \label{fig:synim_pipeline}
\end{figure}

Synthetic calibration tools could apply these transformations sequentially, interpolating the high-resolution phase screens at each step. This sequential resampling acts as an iterative spatial low-pass filter, degrading the high-frequency content of the influence functions. SynIM solves this by implementing a \textit{Combined} workflow based on composite affine transformations. The transformation matrices of the DM (shift, rotation, magnification) and the WFS (including anamorphosis) are mathematically multiplied into a single composite matrix. The high-resolution grid is then transformed and interpolated exactly once, faithfully preserving the spatial frequency content (as visually summarized in Figure \ref{fig:synim_pipeline}).

\subsection{Slope Computation: Numerical Derivatives vs. Telescoping Sum}
\label{subsec:slope_computation}

Extracting WFS slopes from phase screens is a critical step in synthetic calibration. A historical challenge in model-based AO is accurately simulating the Shack-Hartmann centroid displacement on a discrete computational grid. 

A common approach estimates the Gradient-tilt (G-tilt) by computing the discrete local gradient at every pixel using finite central differences, followed by a spatial average over the subaperture. However, this method is intrinsically sensitive to spatial quantization and sub-pixel misalignments that naturally occur during the resampling and downsampling of discrete high-resolution grids. Because of this high demand for geometric exactness, analytical approaches like the boundary-driven telescoping sum—which evaluates the phase difference strictly at the subaperture edges—are structurally more robust. As long as a geometric projection engine operates within standard discretization limits, the telescoping sum effectively shields the slope estimation from internal fractional-pixel artifacts, traditionally making it the safer and more reliable choice.

Nevertheless, finalizing the SynIM geometric engine to achieve absolute sub-pixel perfection revealed an unexpected result. By enforcing an exact spatial mapping—specifically, applying a bilinear upscaling to the nearest exact multiple of the WFS grid prior to reduction—SynIM virtually eliminates asymmetric truncation and grid-shift artifacts. We demonstrated that once this extreme level of geometric accuracy is guaranteed, the numerical derivative method reaches its full accuracy, closely mimicking the physical behavior of a real optical Shack-Hartmann sensor. Consequently, the geometrically aligned numerical derivative method outperforms the traditional analytical estimation of the telescoping sum, establishing itself as the most physically accurate and computationally efficient engine for slope extraction.

\subsection{Overcoming Interpolation Artifacts: The Edge Repair Algorithm}
\label{subsec:edge_repair}

A subtle but critical issue in synthetic calibration is the corruption of phase values at the pupil or subaperture boundaries due to bilinear interpolation during the affine transformation. Interpolation unavoidably ``softens'' the binary mask edges, causing the physical phase values to artificially collapse towards zero. If left untreated, these corrupted edge pixels severely distort the derivative computation, creating artificial high-frequency artifacts in the Interaction Matrix.

\begin{figure}[ht]
    \centering
    \includegraphics[width=0.9\linewidth]{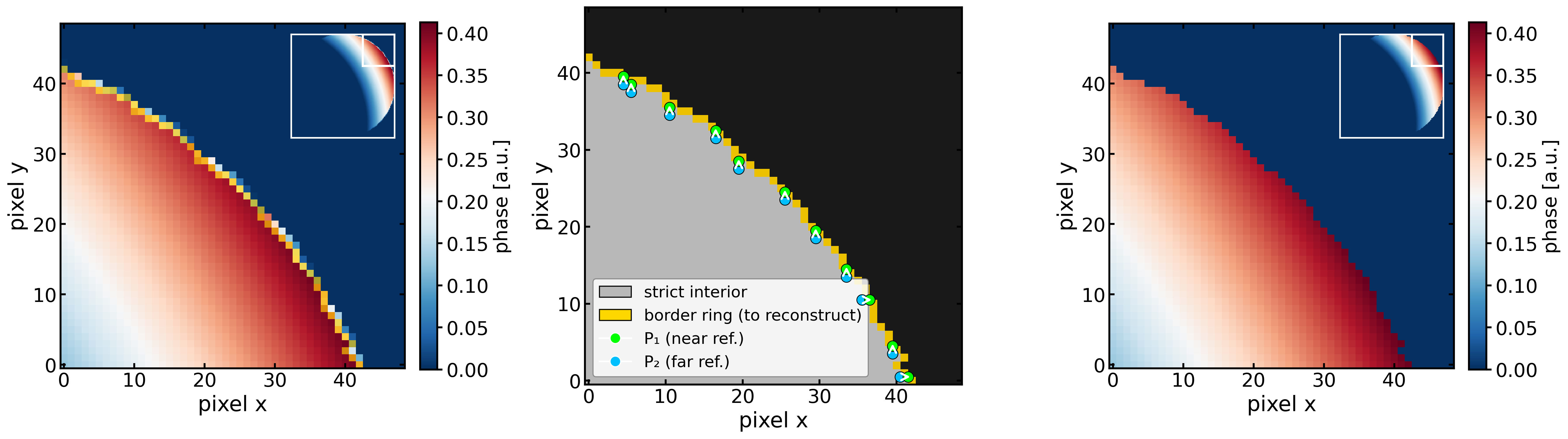}
    \caption{The Edge Repair Algorithm. The corrupted interpolation drop-off at the pupil boundaries is identified (yellow border ring) and perfectly restored using omnidirectional extrapolation from uncorrupted reference pixels ($P_1, P_2$) strictly inside the valid core.}
    \label{fig:edge_repair}
\end{figure}

To solve this, SynIM implements a rigorous edge-repair algorithm (\texttt{\_repair\_interpolated\_phase}). The algorithm operates in three steps (illustrated in Figure \ref{fig:edge_repair}):
\begin{enumerate}
    \item \textbf{Strict Binarization and Clearing:} A stringent threshold (e.g., $0.999999$) is applied to the interpolated mask to isolate the ``pure'' core of uncorrupted phase pixels. The phase data outside this core is intentionally zeroed out to effectively remove the corrupted interpolation drop-off.
    \item \textbf{Edge Identification:} Morphological binary dilation is applied to the strict mask to identify the exact ring of adjacent edge pixels that require restoration.
    \item \textbf{Omnidirectional Extrapolation:} For each identified edge pixel, the algorithm probes all four cardinal directions to locate valid, consecutive reference pixels ($P_1$ and $P_2$) within the pure core. A boundary estimate is computed for each valid direction via linear extrapolation ($2P_1 - P_2$), or constant extrapolation ($P_1$) if only one reference pixel is available. Crucially, to ensure a smooth and artifact-free boundary, the final restored phase value is computed as the normalized average of the extrapolations from all valid directions.
\end{enumerate}
This mitigates boundary collapse, allowing the engine to compute the correct wavefront gradient right up to the true edge of the optical pupil.

\subsection{Tomographic Operators and MMSE Reconstruction}
\label{subsec:tomography}

Once the high-fidelity interaction matrices are computed, SynIM provides built-in modules to assemble tomographic operators. The generation of the Minimum Mean Square Error (MMSE) reconstructors relies on optimal wavefront reconstruction strategies \cite{Fusco2001}. 

SynIM generates the atmospheric covariance matrices ($C_{atm}$) by projecting the expected $C_n^2$ turbulence profile onto the defined modal bases, applying analytical formulations for von Kármán turbulence (accounting for $r_0$ and $L_0$). Similarly, the noise covariance matrices ($C_{noise}$) are constructed directly from detector parameters (readout noise, photon flux). For ELT-scale LGS systems, SynIM natively incorporates analytical spot elongation models, adjusting the noise covariance dynamically based on the sodium layer thickness, laser launch telescope position, and zenith angle~\cite{2010JOSAA..27A...1B}. 

\subsection{Tomographic Projection Matrices and Computational Optimization}
\label{subsec:projection}

In Multi-Conjugate Adaptive Optics, once the atmospheric volume is reconstructed on virtual layers, the wavefront must be projected onto the physical DMs to optimize correction over specific scientific fields of view. Following the optimal formulation by Fusco et al. \cite{Fusco2001}, SynIM computes the tomographic projection matrix $P_{opt}$ by evaluating a pseudo-inverse of the form $(P_{DM}^T P_{DM} + \lambda I)^{-1} P_{DM}^T P_{Layer}$, where $P_{DM}$ and $P_{Layer}$ map the influence functions over the optimized directions.

While the generation of this matrix is straightforward mathematically, it poses a significant computational bottleneck. The computation is heavily dominated by large, dense matrix-matrix multiplications executed over the high-resolution grid. Currently, the spatial sampling of these matrices in SynIM is dictated by the requirements of the SPECULA atmospheric simulation, which typically demands a resolution of 5 to 10 cm per pixel. 

However, there is considerable margin for optimization. In a system like MORFEO, the minimum effective pitch of the DMs—even when considering advanced super-resolution control strategies \cite{2022A&A...667A..48O, 2025ExA....60...26C}—remains between 25 and 50 cm, so operating the projection algorithm at a 5 cm resolution is unnecessarily dense. Introducing a $2 \times 2$ or larger spatial binning prior to the matrix multiplications is therefore a clear optimization path, with negligible impact on the precision of the final projection matrix, deferred to future releases.

\section{VALIDATION AGAINST DIFFRACTIVE MODELS}
\label{sec:validation}

To validate the physical fidelity of the synthetic matrices, we compared the outputs of SynIM against interaction matrices generated using a full diffractive physical optics propagation model via the PASSATA framework\footnote{PASSATA served as the official simulation framework for MORFEO until the Final Design Review (FDR). While SPECULA has since become the current standard, the diffractive results from both frameworks are numerically equivalent for the purposes of this validation.}. We selected the MORFEO (ELT) LGS configuration as our reference test case. This configuration represents a highly demanding scenario: it includes off-axis sources, mismatched relative rotations between the DMs and WFSs (optimized to align the LGS spot elongation with the subaperture diagonals), and altitude-dependent magnification due to the cone effect.

\begin{figure}[ht]
    \centering
    \includegraphics[width=0.6\linewidth]{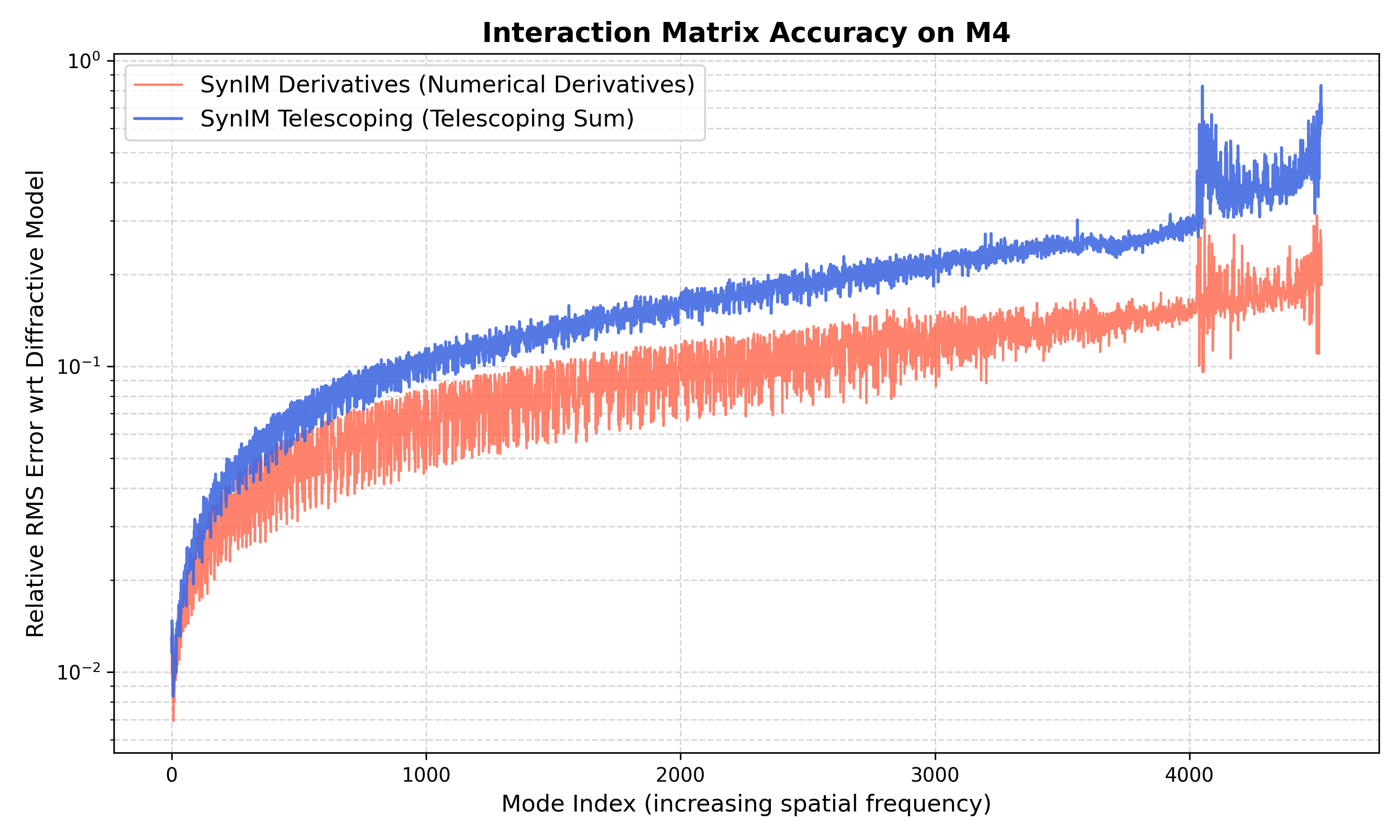}
    \caption{Modal relative RMS error between the SynIM synthetic matrices and the diffractive interaction matrix for the ELT M4 (ground-conjugated DM). With proper sub-pixel geometric alignment enforced, the numerical derivative method (red) outperforms the telescoping sum estimation (blue) across all spatial frequencies.}
    \label{fig:plot1_relative_error_DM0}
\end{figure}
\begin{figure}[ht]
    \centering
    \includegraphics[width=0.6\linewidth]{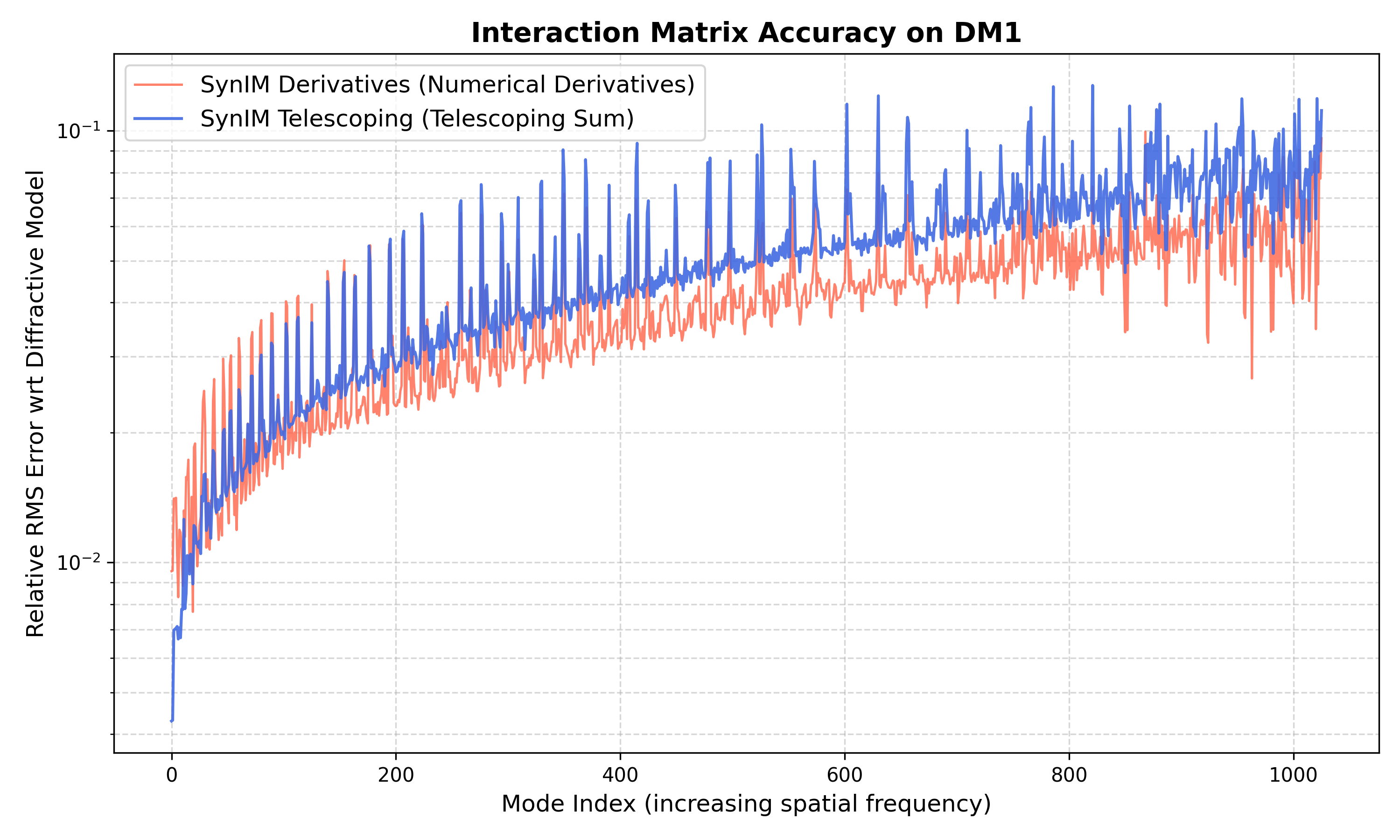}
    \caption{Modal relative RMS error for the MORFEO high-altitude DM (DM1). The periodic peaks typical of high-frequency aliasing at the meta-pupil edges heavily impact the rigid boundary-driven telescoping sum, while the numerical derivatives exhibit remarkable resilience and superior fidelity.}
    \label{fig:plot1_relative_error_DM2}
\end{figure}

Figures \ref{fig:plot1_relative_error_DM0} and \ref{fig:plot1_relative_error_DM2} illustrate the relative RMS error of the synthetic matrices with respect to the diffractive baseline, plotted as a function of the modal index. In the ELT M4 scenario (Figure \ref{fig:plot1_relative_error_DM0}), the laser guide star footprint covers only approximately 90\% of the pupil, and the wavefront is sampled by a $68 \times 68$ subaperture Shack-Hartmann sensor. 

The results highlight an interesting nuance in slope estimation. The analytical G-tilt estimation via telescoping sum (blue) remains a mathematically exact and highly robust estimator of the pure geometric gradient. However, real physical wavefront sensors exhibit diffractive effects and optical cross-talk that naturally smooth these rigid geometric boundaries. The SynIM numerical derivative engine (red), empowered by the strict sub-pixel geometric alignment detailed in Section \ref{subsec:slope_computation}, seamlessly absorbs these continuous optical effects, tracking the diffractive centroid response with a remarkably low relative RMS error ($\sim 10^{-2}$).

A similar behavior is observed for the high-altitude MORFEO DM (Figure \ref{fig:plot1_relative_error_DM2}). High-altitude beams diverge and sweep across different zones of the DM, generating high-frequency aliasing at the meta-pupil edges. Because the telescoping sum relies strictly on exact boundary values, it is naturally more sensitive to these localized geometric complexities, leading to periodic variations in the estimation. In contrast, by continuously averaging the internal footprint, the numerical derivative method naturally mitigates these high-frequency boundary effects, confirming it as the most effective computational engine for capturing the physical response of ELT-scale matrices.

Ultimately, the definitive validation of any synthetic calibration tool resides in its end-to-end closed-loop performance. To verify this, we deployed the SynIM-generated interaction matrices within a full MORFEO MCAO end-to-end simulation. The results confirm that the MMSE reconstructors built using the geometrically-aligned numerical derivative engine deliver an Adaptive Optics correction performance practically equivalent to those derived from the computationally heavy PASSATA physical optics matrices.
Specifically, for the on-axis direction, over a total residual wavefront error budget of $\sim 250$ nm RMS (median atmospheric conditions, bright NGSs, and excluding instrumental and environmental disturbances such as telescope errors, wind shake, phasing errors, or relay aberrations), the performance difference was found to be sub-nanometric, with SynIM achieving $252.6$ nm RMS compared to $252.9$ nm RMS for the PASSATA baseline. Equivalent agreement, with differences below 1 nm RMS, is found across off-axis directions over the scientific and technical field of view. This proves that SynIM not only accurately models theoretical diffractive responses but translates this fidelity seamlessly into state-of-the-art closed-loop performance.

\section{PERFORMANCE AND BENCHMARKING}
\label{sec:performance}

To demonstrate the computational efficiency and scalability of SynIM, we benchmarked the execution times and memory footprint using the full baseline MORFEO ELT configuration. The tests were executed on a computational node equipped with a 48-core AMD Epyc 9454 CPU, local Solid-State Drive (SSD) storage, and a single NVIDIA L40S GPU with 46 GB of VRAM. To optimize memory bandwidth while preserving numerical stability, all computations were performed in single precision (\texttt{float32}). 

\subsection{GPU Acceleration: CPU vs. GPU computation}
Generating a synthetic Interaction Matrix for an ELT-scale ground-layer deformable mirror (such as M4, controlling 4519 modes over a high-resolution grid) requires massive array manipulations and interpolations. When executed purely on the CPU via standard NumPy/SciPy parallelized backends, the generation of a single LGS interaction matrix using the standard algorithm requires 13.1 seconds. 

By leveraging the CuPy backend, SynIM executes the identical pipeline on the GPU in just 1.0 seconds. This represents a $13\times$ speed-up, confirming that hardware acceleration effectively shifts synthetic calibration from a slow offline task to an operation compatible with daytime calibration constraints. During this operation, the peak GPU memory footprint reached 41.6 GB. While the base single-precision 3D influence function cube for 4519 modes accounts for roughly 4.5 GB, the execution of composite 3D affine transformations requires the allocation of temporary intermediate arrays and the final output tensor, which momentarily elevates the VRAM footprint during the interpolation step.

\subsection{Multi-WFS Batch Optimization and Tomographic Generation}
MORFEO employs a split-tomography architecture featuring 12 WFSs. To generate the complete tomographic control dictionary for the LGS channel, interaction matrices must be computed across 12 independent altitude components: 3 physical DMs and 9 virtual atmospheric layers, mapping altitudes from 0 to 22 km. Due to the diverging optical beams, the computational grids for these layers scale dynamically from $480 \times 480$ up to $690 \times 690$ pixels, encompassing over 41,000 active modes in total. Computing this volume for 6 LGS WFSs yields 72 distinct high-resolution matrices.

In a naive sequential implementation, generating these matrices would require reading and transferring massive influence function cubes repeatedly. SynIM optimizes this via its \textit{Multi-WFS} batch module. When multiple sensors observe the same components from different guide star coordinates, SynIM fetches the high-resolution influence function cube exactly once per altitude layer. Subsequently, only the WFS-specific geometric projections and the dense computations are iterated over the 6 sensors on the GPU.

Benchmarking this full-scale tomographic generation shows that SynIM computes the entire set of 72 ELT-class interaction matrices using the numerical derivative engine in 265.6 seconds. This corresponds to an average processing time of 3.7 seconds per matrix, which inherently includes the increased computational load associated with larger meta-pupil grids at high altitudes.

Profiling the execution indicates that the overall time is driven by the execution of dense 3D affine transformations on the GPU, as well as the transfer of approximately 50 GB of influence function data across the PCIe bus. Ultimately, combining sub-pixel geometrical alignment with vectorized numerical derivatives provides a viable and scalable solution to meet the computational requirements of ELT operations, RTC updates, and repetitive Monte Carlo design simulations.

\section{INTEGRATION WITH ONLINE CONTROL AND NEXT-GENERATION FACILITIES}
\label{sec:applications}

While initially conceived as an offline tool for system design and initial calibration, the computational efficiency of SynIM opens new avenues for online AO control. Next-generation systems suffer from slowly varying optical misregistrations (e.g., pupil shifts, rotations) and changing atmospheric conditions during observations. 

To address this, the integration of SynIM with SPRINT (System Parameters Recurrent INvasive Tracking) \cite{2021MNRAS.504.4274H, 2024SPIE13097E..5PA} has been implemented and successfully demonstrated within the SPECULA simulation framework. In this simulated architecture, the SPRINT estimator continuously demodulates the WFS slopes to extract the measured interaction matrix from closed-loop telemetry. SynIM acts as the high-speed backend engine: it ingests the current parameter estimates and leverages its GPU-accelerated \textit{Combined} affine transformations to rapidly generate updated synthetic interaction matrices and MMSE reconstructors on-the-fly. Although substantial integration work remains before actual on-sky commissioning, these simulations validate the feasibility of the envisioned workflow: an iterative refinement loop that updates the Real-Time Computer (RTC) models without interrupting scientific observations.

Beyond the baseline MORFEO configuration, the flexibility of the SynIM architecture—natively supporting SCAO, MCAO, LTAO, and GLAO via simple YAML configurations—makes it a highly suitable calibration engine for a wide range of future and upgraded facilities. It possesses the necessary features and computational scalability to potentially support the demanding visible-wavelength MCAO calibration for MAVIS (VLT), advanced LTAO architectures like KAPA (Keck), operations for the AOF (VLT), and the extreme wide-field requirements of the Wide Field Spectroscopic Telescope (WST).

\section{CONCLUSIONS}
\label{sec:conclusions}

The transition from empirical to Model-Based calibration is a strict requirement for the success of next-generation telescopes, particularly Extremely Large Telescopes operating with thousands of degrees of freedom. In this paper, we presented SynIM, a highly optimized, GPU-accelerated Python library designed to address the severe numerical and computational challenges inherent to synthetic calibration at this extreme scale. 

By merging all geometric transformations into a single composite operation, and pairing strict sub-pixel grid alignment with a rigorous edge-repair algorithm, SynIM successfully eliminates cumulative interpolation and boundary truncation artifacts.
This geometric rigor revealed that standard numerical derivatives, which are heavily penalized by discrete grid misalignments in naive implementations, actually provide a highly faithful representation of real Shack-Hartmann sensors. Validated against the rigorous PASSATA framework on ELT-scale configurations, SynIM's optimized numerical derivative engine demonstrated superior accuracy and resilience to meta-pupil boundary effects compared to geometric estimators like the telescoping sum, while computing a complete MCAO dictionary of 72 matrices in under 5 minutes.

Furthermore, end-to-end simulations of the full MORFEO system verify that SynIM-generated matrices yield closed-loop correction performance practically equivalent to those obtained via computationally heavy physical optics models, demonstrating the operational readiness of the library. Seamlessly integrated with modern simulation environments like SPECULA and primed to act as the matrix-generation engine for online tracking strategies like SPRINT, SynIM stands as a versatile, high-fidelity calibration tool ready to support the next era of astronomical adaptive optics.

\acknowledgments % equivalent to \section*{ACKNOWLEDGMENTS}       
 
The authors would like to acknowledge that the development of the SynIM library was primarily carried out within the context of the MORFEO (Multi-conjugate adaptive Optics Relay For ELT Observations) project. We extend our gratitude to the MORFEO consortium and the ADaptive Optics National laboratory in Italy (ADONI) for their continued support, the fruitful discussions, and the stimulating scientific environment that made this work possible.

% References
\bibliography{report} % bibliography data in report.bib
\bibliographystyle{spiebib} % makes bibtex use spiebib.bst

\end{document}